# Accurate Electronic, Transport, and Bulk Properties of Gallium Arsenide (GaAs)

Yacouba Issa DIAKITE[1], Sibiri D. TRAORE[1], Yuriy Malozovsky[2], Bethuel Khamala[3], Lashounda Franklin[2], and Diola BAGAYOKO[2]

[1]Departement of Studies and Research (DSR) in Physics, College of Science and Technology (CST), University of Science, Techniques, and Technologies of Bamako (USTTB), Bamako, Mali

[2]Department of Mathematics, Physics, and Science and Mathematics Education (MP-SMED), Southern University and A&M College in Baton Rouge (SUBR), Louisiana, USA

[3] Department of Computational Sciences, University of Texas at El Paso (UTEP), El Paso, Texas, USA

## Abstract

We report accurate, calculated electronic, transport, and bulk properties of zinc blende gallium arsenide (GaAs). Our ab-initio, non-relativistic, self-consistent calculations employed a local density approximation (LDA) potential and the linear combination of atomic orbital (LCAO) formalism. We strictly followed the Bagayoko, Zhao, and William (BZW) method, as enhanced by Ekuma and Franklin (BZW-EF). Our calculated, direct band gap of 1.429 eV, at an experimental lattice constant of 5.65325 Å, is in excellent agreement with the experimental values. The calculated, total density of states data reproduced several experimentally determined peaks. We have predicted an equilibrium lattice constant, a bulk modulus, and a low temperature band gap of 5.632 Å, 75.49 GPa, and 1.520 eV, respectively. The latter two are in excellent agreement with corresponding, experimental values of 75.5 GPa (74.7 GPa) and 1.519 eV, respectively. This work underscores the capability of the local density approximation (LDA) to describe and to predict accurately properties of semiconductors, provided the calculations adhere to the conditions of validity of DFT.[16]

## I.    Introduction

Gallium arsenide is an important electronic and opto-electronic material.[1] It is a prototypical binary semiconductor. It has  a high electron mobility and a small dielectric constant; GaAs is extensively utilized in  high temperature resistance, ultrahigh frequency, low-power devices and circuits.[2] Gallium arsenide crystallizes in zinc-blende structure; many  experiments and theoretical works established that it has a direct band gap. Several experimental reports dealt with the room temperature band gap of the material. Room temperature band gaps as small as 1.2 eV[3]  and as high as to 1.7 eV[4] have been reported.  Dong et al.[4] attributed the significant difference between these two values  to a tip-induced band bending in the semiconductor.

Recent, experimental values of the room temperature band gap of GaAs are 1.42 eV[5], 1.425 eV[6] and 1.43 eV.[7] The accepted value of the room temperature band gap is 1.42 eV[5] to 1.43 eV[7]; These value are in basic agreement with 1.425 eV and 1.430 eV. In the bottom rows of Table I, we show over 10 different measurements of the band gap of GaAs. As per the content of this table, the consensus experimental band gap, at low temperature, is 1.519 eV.[8-10]

Numerous theoretical results have been reported for the band gap of GaAs. Our focus on the band gap stems from its importance in describing several other properties of semiconductors [AIP Advances]; in particular, a wrong bang gap precludes agreements between peaks in the calculated densities of states, dielectric functions, and optical transition energies with their experimental counterparts. In contrast to the consensus reached for the room and low temperature experimental gaps for GaAs, the picture for theoretical results is far from being satisfactory. Indeed, numerous theoretical values of the band gaps, obtained from ab-initio calculations, disagree with each other and disagree with experiment. Table I contains over 28 band gaps calculated with a local density approximation (LDA) potential. Some 16 of these results, from ab-initio calculations, range from 0.09 eV[11, 12] to 0.98 eV.[13]

Other results obtained with LDA potentials, as shown in Table I, are either underestimates or overestimates of the band gap of GaAs, except for some three that require some comments. The linear muffin tin orbital (LMTO) calculation that obtained a gap of 1.46 eV[14] employed an additional potential besides the standard LDA.  The ab-initio LDA calculation that obtained a band gap of 1.54 eV[15] employed a lattice constant of 5.45 Å, a value that is 3% smaller than the low temperature value in Table I.  As explained elsewhere,[16] the Tran and Blaha  modified Becky and Johnson potential (TB-mBJ)[17] is not entirely a density functional one – given that it cannot be obtained from the functional derivative of an exchange correlation energy functional.[16,18]  So, while two calculations with this potential led to gaps of 1.46 eV[19] and 1.56 eV,[20] in general agreement with experiment, these values do not resolve the woeful underestimation by most of the LDA and GGA calculations in Table I.

As shown in Table I, 12 calculations employing a generalized gradient approximation (GGA) found band gap values varying from 0.206 eV[19] to 1.03 eV.[21] Only one GGA calculation found a gap of 1.419 eV,[22] in basic agreement with the above accepted, experimental gaps of 1.42 eV – 1.43 eV and 1.519 eV for room and low temperatures, respectively. The calculation that utilized a meta-GGA potential found a gap of 1.276 eV,[19] smaller than the experimental one.

The Green function and dressed Coulomb (GW) approximation calculations led to mixed results. The non-self-consistent $G_0W_0$ calculation obtained a gap of 1.51 eV,[23] in agreement with the low temperature experimental value of 1.519 eV, while the self-consistent GW calculation produced 1.133 eV,[24] well below the low temperature value. Several other theoretical results are reported in Table I. Some utilized a hybrid functional potential,[23] while others employed the modified Becke and Johnson (mBJ) potential.[16] These potentials are different from the standard, ab-initio LDA or GGA potentials due to the utilization of one or more parameters in their construction.

The results of calculations employing these potentials vary with those parameters. For this reason, these results, while very useful, do not resolve the fundamental question of the serious band gap underestimation. With the use of several fitting parameters, the three empirical pseudo potential calculations, shown in Table I, understandably led to the correct, low temperature experimental band gap of GaAs.

The above overview of the literature points to the need for our work. Indeed, numerous calculated values of the band gap disagree with corresponding, experimental ones. The disagreement between sets of calculated band gaps, as evident above and in Table I, adds to our motivation for this work. At the onset, we have to answer the question as to the reason our LDA calculations can be expected to lead to an accurate description of electronic and related properties of GaAs. Past, accurate descriptions[16] and predictions[16] of properties of semiconductors, using the distinctive feature our calculations, portend the same for GaAs. This distinctive feature, the Bagayoko, Zhao, and Williams (BZW) method, as enhanced by Ekuma and Franklin (BZW-EF), strictly adheres to conditions of validity of DFT or LDA potentials, as elucidated by Bagayoko.[16]

We are aware of some explanations of the failures of many previous calculations to lead to correct values of the band gaps of semiconductors or insulators. Prominent among them are the self- interaction (SI)[25] and the derivative discontinuity[26-28] of the exchange correlation energy. Bagayoko,[16] using strictly DFT theorems and the Rayleigh theorem for eigenvalues, demonstrated that self-consistent calculations that do not adhere to well-defined, intrinsic features of DFT cannot claim to produce eigvenvalues and other quantities that possess the full, physical content of DFT. Hence, disagreements between their results and experiment may arise mostly from the fact that their findings do not fully possess the physical content of DFT. Our perusal of the articles that reported the results in Table I did not lead to any publication that adhered totally to these features of DFT. Specifically, we could not find any calculation that methodically searched for and attained the absolute minima of the occupied energies, using increasingly larger and embedded basis sets.[16] The point here is that popular explanations of band gap underestimation by DFT calculations notwithstanding, our distinctive computational method is likely to describe GaAs accurately.

The rest of this paper is organized as follows. This section, devoted to the introduction, is followed by a description of our computational method, in Section II. We subsequently present our results in Section III and discuss them in Section IV. Section V provides a short conclusion.

**Table I.** Calculated, fundamental bang gaps (Eg, in eV) of zinc blend GaAs, along with pertinent lattice constants in Angstroms, and experimental values.

| Computational Formalism | Potentials (DFT and others) | a(Å) | Eg(eV) |
|---|---|---|---|
| LMTO | LDA (fully relativistic local density) | | 0.25[a] |
| | LDA+$V_w$(with extra potentials) | | 1.46[a] |
| LCGO | LD | 5.654 | 1.21[b] |
| LAPW | LDA | 5.653 | 0.28[c] |
| PAW | LDA | | 0.330[d] |
| FP -LAPW | LDA | 5.6079 | 0.463[e] |
| | LDA | | 0.28[f] |
| Self-consistent DFT | LDA-SZ | 5.68 | 0.61[g] |
| | LDA-SZ-O | 5.66 | 0.78[g] |
| | LDA-SZP-O | 5.60 | 0.98[g] |
| | LDA-DZ | 5.64 | 0.66[g] |
| | LDA-DZP | 5.60 | 0.82[g] |
| | LDA-PW | 5.55 | 1.08[h] |
| | LDA-PW | 5.55 | 0.7[i] |
| First-principal total-energy calculations | LDA | | 1.17[j] |
| First-principal total-energy calculations | LDA | | 1.23[k] |
| Plane-wave pseudopotential | LDA | 5.45 | 1.54[l] |
| Plane-wave pseudopotential | LDA | 5.654 | 1.04[m] |
| FP-LAPW | LDA | | 1.613[n] |
| UPP (CASTEP) | LDA-mBJ | | 1.46[n] |
| | LDA-sX | | 1.639[n] |
| FP-LAPW NCP (SIESTA) | LDA | | 0.54755[n] |
| | LDA | | 0.23[o] |
| | LDA | | 0.18[p] |
| | LDA | | 0.09[q] |
| | LDA | | 0.32[r] |
| | GGA | | 0.51[f] |
| | GGA-EV | | 1.03[f] |
| | GGA-EV | | 0.97[s] |
| | GGA | | 0.49[r] |
| FP-LAPW | GGA-PBE | | 0.329[n] |
| UPP (CASTEP) | GGA-WC | | 0.206[n] |
| | GGA-PBE | | 0.52317[n] |
| FP-LAPW NCP (SIESTA) | Meta-GGA | | 1.27637[n] |
| Ab initio pseudopotential | GGA | 5.653 | 1.419[t] |
| All electron atomic orbit | GGA | | 0.82[u] |
| PAW | GGA | 5.734 | 0.674[v] |
| PAW | GGA-PBE | 5.648 | 0.43[w] |
| FP -LAPW | GGA-WC | 5.6654 | 0.341[e] |
| | GGA-EV | | 0.968[e] |
| | mBJ-LDA | | 1.560[e] |
| | mBJ-LDA | | 1.64[x] |
| | HSE06 | | 1.33[w] |
| | $G_0W_0$ | | 1.51[w] |
| Plane wave and pseudopotential | GW | | 1.133[d] |
| | SX | | 1.289[d] |
| Ab initio pseudopotentials | | 5.52 | 0.4[y] |
| LUC-INDO | | 5.6542 | 1.91[z] |
| EPM | | | 1.527[α] |
| EPM | Non local pseudopotential | | 1.51[β] |
| EPM | Non local pseudopotential | 5.65 | 1.51[γ] |

**Experiments**

| | | | |
|---|---|---|---|
| Experimental | Absorption spectra measurements | | 1.519[δ] at low T |
| | Photoluminescence measurements | | 1.519[ε], low T |
| | | | 1.43[ζ] at 300 K |
| | Magnetoluminescence measurements | 5.65325 | 1.5192[η], low T |
| | Transmission measurements | | 1.42[θ] at 300 K |
| | Raman measurements | | 1.519[ι] at low T |
| | | | 1.425[ι], 300 K |
| | Scanning tunneling microscopy and spectroscopy measurements | | 1.7[κ] at 300 K |
| | | | 1.2[λ] at 300 K |
| | | | 1.42[μ] at 300K |
| | Photocapacitance measurements | | 1.5[ν] at 77 K |

[a]Ref[14], [b]Ref[29], [c]Ref[30], [d]Ref[24], [e]Ref[20], [f]Ref[21], [g]Ref[13], [h]Ref[31], [i]Ref[32], [j]Ref[33], [k]Ref[34], [l]Ref[15], [m]Ref[35], [n]Ref[19], [o]Ref[36, 37], [p]Ref[38, 39], [q]Ref[11, 12], [r]Ref [40, 41] , [s]Ref[42], [t]Ref[22], [u]Ref[43], [v]Ref[2], [w]Ref[23], [x]Ref[44], [y]Ref[45], [z]Ref[46], [α]Ref[47], [β]Ref[48], [γ]Ref[49], [δ]Ref[8], [ε]Ref[9], [ζ]Ref[7], [η]Ref[10], [θ]Ref[50], [ι]Ref[6], [κ]Ref[4], [λ]Ref[3], [μ]Ref[5], [ν]Ref[51].

## II.    Computational Approach and the BZW-EF Method

Our calculations are similar to most of the previous ones discussed in Table I, as far as the choice of the potential and the use of the linear combination of atomic orbitals (LCAO) are concerned. We used the local density approximation (LDA) potential of Ceperley and Alder[52] as parameterized by Vosko, Wilk and Nusair.[53] We employed Gaussian functions in the radial parts of the atomic orbitals, resulting in the linear combination of Gaussian orbitals (LCGO). The distinctive feature of our calculations, as compared to the ones discussed above, stems from our implementation of the LCGO formalism following the Bagayoko, Zhao, and Williams (BZW) method, as enhanced by Ekuma and Franklin (BZW-EF).[16, 54-55]

The method searches for the absolute minima of the occupied energies, using successively augmented basis sets, and avoids the destruction of the physical content of the low, unoccupied energies – once the referenced minima are attained. Typically, the implementation starts with a self-consistent calculation that employs a small basis set; this basis set is not to be smaller than the minimum basis set, the one that can just account for all the electrons in the system. A second calculation follows, with a basis set consisting of the previous one plus one additional orbital. The dimension of the Hamiltonian matrix is consequently increased by 2, 6, 10, or 14 for s, p, d, and f orbitals, respectively. Upon the attainment of self-consistency, the occupied energies of Calculation II are compared to those of I, graphically and numerically. In general, upon setting the Fermi level to zero, some occupied energies from Calculation II are found to be lower than corresponding ones from Calculation I. This process of augmenting the basis set and of comparing the occupied energies of a calculation to those of the one immediately preceding it continues until three consecutive calculations lead to the same occupied energies. This criterion is a clear indication of the attainment of the absolute minima of the occupied energies. The first of these three calculations, with the smallest basis set, is the one that provides the DFT description of the material. The basis set for this calculation is the optimal basis set.

While the second of these calculations generally leads to the same occupied and low, unoccupied energies up to 6-10 eV, depending on the material, the third of these calculations often lowers some low, unoccupied energies from their values obtained with the optimal basis set. We should note that the referenced three calculations lead to the same electronic charge density. As explained by Bagayoko,[16] the energy functional derived from the Hamiltonian is a unique functional of the ground state charge density. Hence, the occupied and unoccupied energies of the spectrum of this Hamiltonian, *with the physical content of DFT*, cannot change upon an increase of the basis set. Consequently, the unoccupied energies obtained with basis sets much larger than the optimal basis set, and that contain this set, do not represent DFT solutions if they differ from their corresponding values obtained with the optimal basis set.

Bagayoko[16] explained the unphysical nature of unoccupied energies, lowered from their values obtained with the optimal basis set, in terms of mathematical artifacts stemming from the Rayleigh theorem for eigenvalues. Upon the attainment of the absolute minima of the occupied energies, the above extra lowering of some unoccupied energies, with increasing basis sets, is not only a possible explanation of the underestimation of band gaps by calculations that do not search and find the optimal basis set, but also of discrepancies between several calculations that utilize the same potential and computational formalism as shown in Table I.

The following computational details are intended to facilitate the replication of our work. GaAs is III-V semiconductor, with the zinc blende crystal structure in normal conditions of temperature and pressure. We used the experimental, room temperature lattice constant of 5.65325 Å.[56] Ab-initio calculations of the electronic structures of $Ga^{+1}$ and $As^{-1}$ produced atomic orbitals employed in the solid state calculation. We utilized even-tempered Gaussian exponents, with 0.28 as the minimum and $0.55 \times 10^5$ as the maximum, in atomic unit, for $Ga^{+1}$. We used 18

Gaussian functions for s and p orbitals and 16 for the d orbitals. Similarly, the Gaussian exponents for describing $As^{-1}$ were from 0.2404 to 0.349 x $10^5$. A mesh size of 60 k points in the irreducible Brillouin zone, with appropriate weights, was used in the iterations for self-consistency. The computational error for the valence charge was about 1.25 x $10^{-3}$ per electron. The self-consistent potentials converged to a difference around $10^{-5}$ between two consecutive iterations.

With the LDA potential identified above and the computational details, we implemented the LCGO formalism following the BZW-EF method. Upon the attainment of absolute minima of the occupied energies, the optimal basis set was employed to produce the band structure of GaAs. The resulting eigenvalues and corresponding wave functions were utilized to calculate the total (DOS) and partial (pDOS) densities of states, as well as electron and hole effective masses. From the curve of the calculated total energy versus the lattice constant, we obtained the equilibrium lattice constant and the bulk modulus. These results follow below, in Section III.

### III.    Results

We present below the successive calculations that led to the absolute minima of the occupied energies for GaAs. Then, we discuss the electronic energy bands resulting from the calculation with the optimal basis set. We subsequently show the total (DOS) and partial (pDOS) densities of states and effective masses derived from the energy bands. The last results to be discussed pertain to the total energy curve, the equilibrium lattice constant, and the bulk modulus.  We show, in Table II below, the successive calculations with increasing basis sets, along with the applicable orbitals and calculated band gaps. The occupied energies obtained by Calculations III, IV, and V are identical. Hence, Calculation III provides the DFT description of GaAs.

| Calculation Number | Gallium Orbitals for $Ga^{1+}$ | Orbitals for $As^{1-}$ | No. of Wave Functions | Band Gap in eV |
|---|---|---|---|---|
| Calc. I | $3s^23p^63d^{10}4s^24p^0$ | $3s^23p^63d^{10}4s^24p^4$ | 52 | 1.380 |
| Calc. II | $3s^23p^63d^{10}4s^24p^04d^0$ | $3s^23p^63d^{10}4s^24p^4$ | 62 | 1.368 |
| **Calc. III** | $\mathbf{3s^23p^63d^{10}4s^24p^04d^0}$ | $\mathbf{3s^23p^63d^{10}4s^24p^44d^0}$ | **72** | **1.429** |
| Calc. IV | $3s^23p^63d^{10}4s^24p^04d^05s^0$ | $3s^23p^63d^{10}4s^24p^44d^0$ | 74 | 1.270 |
| Calc. V | $3s^23p^63d^{10}4s^24p^04d^05s^0$ | $3s^23p^63d^{10}4s^24p^44d^05s^0$ | 76 | 1.238 |

The calculated band structure of GaAs, from Calculation III, is shown in Figure 1. As per the explanations provided in the method section, the superposition of the occupied energies from Calculations III, IV, and V signifies that the absolute minima of the occupied energies are reached in Calculation III whose corresponding basis set is the optimal basis set. The calculated, direct band gap at the Γ point is 1.429 eV ($\approx$ 1.43 eV). This value is in excellent agreement with the accepted value for the room temperature experimental band gap of GaAs, i.e., 1.42-1.43 eV. This agreement is in stark contrast with the case of most previous, calculated band gaps in Table I.

Figures 2 and 3 show the total (DOS) and partial (pDOS) densities of states obtained from the bands resulting from Calculation III. Several features of our calculated density of states (DOS) are close or the same as those of experimental densities of states from X-ray photoemission spectroscopy measurements.[57] According to Fig. 14 in the article by Ley et al.,[57] the peak positions of $H_{IT}$, $P_{II}$, and $P_{III}$ correspond to the binding energies of 1.0 eV, 6.6 eV, and 11.4eV, respectively. From our calculations, the corresponding values are 1.0 eV, 6.4 eV, and 11.0 eV, respectively. The labels of the peaks are as reported by Ley et al.[57] As per our calculated pDOS in Fig. 3, the lowest lying group of valence bands is entirely from Ga d, while the middle group consists mostly of As s with faint contributions from Ga s and Ga p. The upper most group of valence bands is clearly dominated by As p, with a significant overlap with Ga s and a smaller contribution from Ga p.

We provide in Table III the calculated, electronic energies between -18 eV and about 10 eV, at high symmetry points in the Brillouin zone. The content of this table is partly intended to enable accurate comparisons with future, experimental measurements from X-ray, ultra violet (UV) or other spectroscopies. From this content, the widths of the upper most, middle, and lower most groups of valence bands are 6.58 eV, 2.494 eV, and 0.203 eV, respectively. The total width of the valence band is 15.903 eV.

**Figure 1**. Calculated, electronic bands of GaAs, as obtained from Calculation III. The calculated, direct band gap, at Γ, is 1.429 eV.

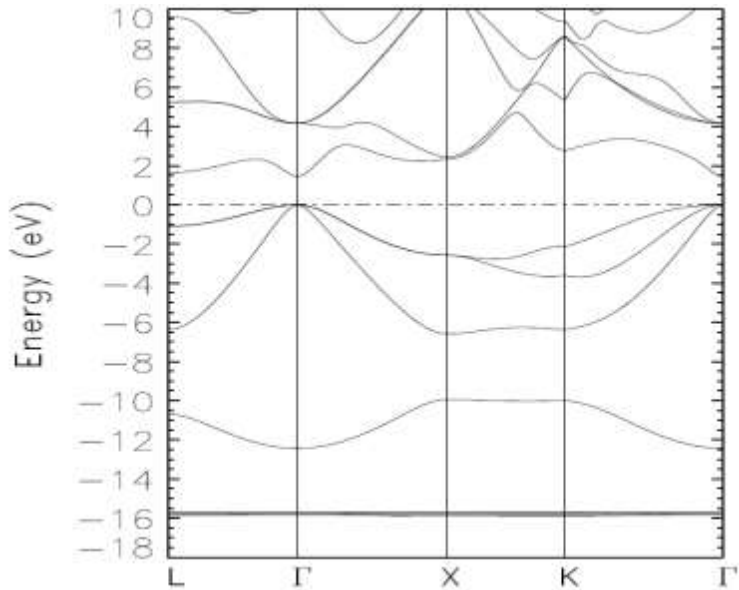

**Figure 2**. Total density of states of GaAs, obtained from the energy bands in Figure 1. The zero on the horizontal axis indicates the position of the Fermi level.

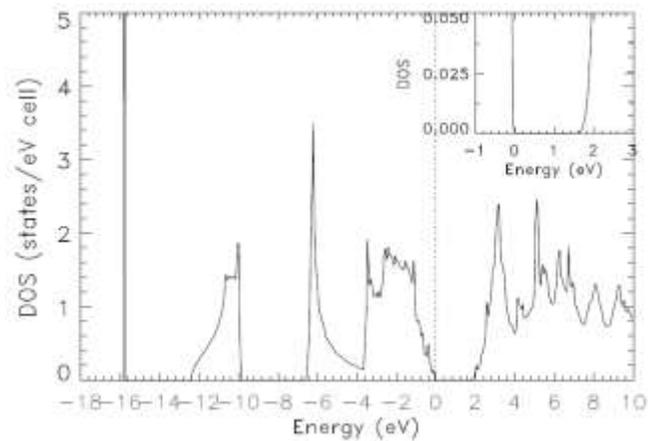

**Figure 3**. Calculated, partial densities of states (pDOS) for GaAs, as derived from the bands from Calculation III, as shown in Fig. 1.

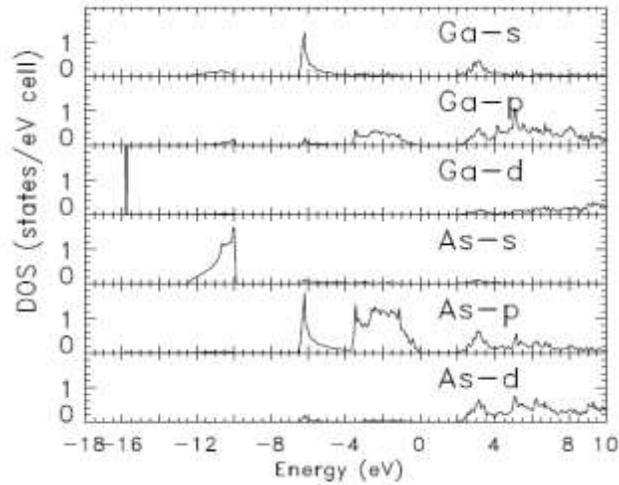

**Table III.** Calculated, electronic energies of GaAs at high symmetry points in the Brillouin Zone, as obtained with the optimal basis set of Calculation III. We used the experimental lattice constant of 5.65325Å. This table is to enable comparisons with future room temperature, experimental and theoretical results.

| L-point | Γ-point | X-point | K-point |
|---------|---------|---------|---------|
| 9.593 | 4.164 | 10.772 | 8.642 |
| 5.248 | 4.164 | 10.772 | 8.590 |
| 5.248 | 4.164 | 2.429 | 5.320 |
| 1.646 | 1.429 | 2.336 | 2.769 |
| -1.095 | 0 | -2.572 | -2.150 |
| -1.095 | 0 | -2.572 | -3.601 |
| -6.370 | 0 | -6.582 | -6.379 |
| -10.697 | -12.439 | -9.945 | -9.987 |
| -15.710 | -15.711 | -15.700 | -15.705 |
| -15.710 | -15.711 | -15.731 | -15.722 |
| -15.802 | -15.821 | -15.778 | -15.783 |
| -15.802 | -15.821 | -15.778 | -15.786 |
| -15.870 | -15.821 | -15.903 | -15.891 |

We calculated the effective masses of n-type carriers for GaAs, using the electronic structure from Calculation III (in Fig.1), i.e., the vicinity of the conduction band minimum at the Γ point. In Table IV, we show our results along with several, previous theoretical and experimental ones. Experimental electron effective masses are directionally averaged. Our results are comparable with those from measurement.

**Table IV**. Calculated, effective masses for GaAs (in units of the free electron-mass, $m_0$) : $m_e$ indicates an electron effective mass at the bottom of the conduction band ; $m_{hh}$ and $m_{lh}$ represent the heavy and light hole effective masses, respectively. Theo = theory, Expt = experiment.

| | **Our Work** | **Theo[a] EPM** | **Theo[b]** | **Theo[c]** | **Theo[d]** | **Expt[e] Room T** | **Expt[f] Room T** |
|---|---|---|---|---|---|---|---|
| $m_e$ ($\Gamma$-L) | 0.077 | 0.066 | 0.012 | 0.070 | | 0.063 | 0.0635 |
| $m_e$ ($\Gamma$-X) | 0.077 | | | | 0.030 | | |
| $m_e$ ($\Gamma$-K) | 0.078 | | | | | | |
| | | | | | | | |
| $m_{hh}$ ($\Gamma$-L) | 0.865 | 0.866 | | 0.827 | | 0.50 | 0.643 |
| $m_{hh}$ ($\Gamma$-X) | 0.359 | 0.342 | | 0.334 | 0.320 | | |
| $m_{hh}$ ($\Gamma$-K) | 0.516 | | | | | | |
| | | | | | | | |
| $m_{lh}$ ($\Gamma$-L) | 0.062 | | | 0.056 | | 0.076 | 0.081 |
| $m_{lh}$ ($\Gamma$-X) | 0.076 | 0.093 | | 0.068 | 0.036 | | |
| $m_{lh}$ ($\Gamma$-K) | 0.070 | | | | | | |

[a]Reference[47],[b]Reference[14],[c]Reference[29],[d]Reference[23], [e]Reference[56],[f]Reference[5]

**Figure 4.** The calculated, total energy of GaAs versus the lattice constant. The minimum total energy is located at 5.632 Å, our predicted equilibrium lattice constant.

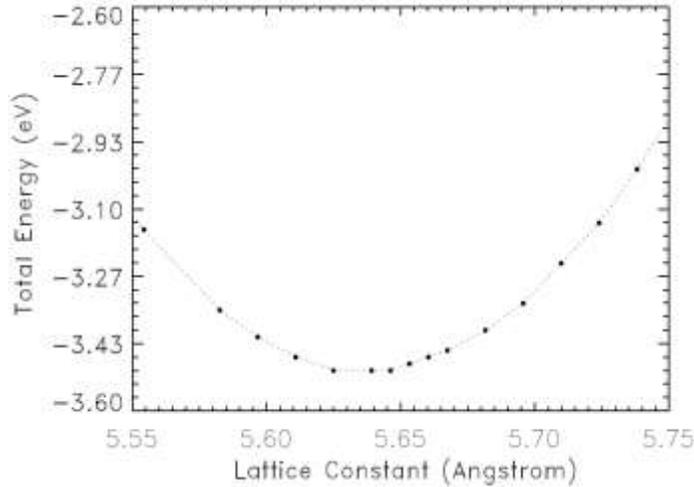

Our calculations predicted the equilibrium lattice constant to be 5.632 Å. The calculated, direct band gap, at the equilibrium lattice constant, is 1.520 eV, at the $\Gamma$ point. This result of 1.520 eV is in excellent agreement with the low temperature experimental value of 1.519 eV, reported by four different, experimental groups.[6, 8-10] Our calculated bulk modulus of 75.49 GPa also agrees with the experimental values of 75.5 and 74.7 GPa.[56, 58]

## IV.    Discussions

From our overview of the literature and the content of Table I, the band gap of GaAs, a prototypical semiconductor, was systematically underestimated by first principle, self-consistent calculations that utilized ab-initio LDA or GGA potentials.  Unlike these previous results, our calculated, direct band gaps of 1.429 eV and 1.520 eV, for room and low temperatures, respectively, are in excellent agreement with corresponding, experimental ones. As shown in the section on results, the locations of several peaks in the calculated, total valence density of states practically agree with corresponding experimental ones. This latter agreement strongly indicates that our calculated band gap values are not fortuitous. Additionally, our calculated effective masses are close to corresponding, available, experimental ones, like some previous, theoretical results. A detailed comparison of the calculated, effective masses with experimental ones is partly hindered by the unavailability of directional, effective masses; most experiments reported averaged values.

Our explanation of the excellent agreements noted above rests on the fact that our calculations, with the BZW-EF method, strictly adhered to necessary conditions[16] for their results to have the physical content of DFT. A careful perusal of the articles reporting the previous results in Table I found no indication that the pertinent calculations searched for and verifiably attained the absolute minima of the occupied energies. *Without this explicit attainment, the results cannot be expected to possess the full physical content of DFT.*[16]   The BZW-EF method invokes the Rayleigh theorem for the selection of the optimal basis set out of several others that lead to the same occupied energies; the smallest of these basis sets, the optimal basis set, is complete for the description of the ground state and is not over-complete, like much larger ones that include it. Different over-complete basis sets containing the optimal one are expected to lead to different underestimated values of the measured band gap.

## V.    Conclusion

We performed ab-initio, self-consistent calculations of electronic, transport, and bulk properties of GaAs. Our results, unlike those of many previous ab-initio calculations, agree very well with experiment, for the band gaps, the total density of states, and the bulk modulus; they also agree with experiment for the effective masses, where the latter are inversely related to the mobility of charge carriers.  We credit our strict adherence to conditions of validity for DFT or LDA potentials, with our implementation of the BZW-EF method, for the above agreements between our calculated results and experimental ones.

**Acknowledgments**

This research was funded by the Malian Ministry of Higher Education and Scientific Research, through the Training of Trainers Program (TTP), the US National Science Foundation [NSF,